# Metallurgical Processes in Al–Si Alloy Improved by WC Nanoparticles.

Konstantin Borodianskiy

Department of Chemical Engineering and Materials, Ariel University, Ariel 40700, Israel

**ABSTRACT**

The influence of a modifier based on WC nanoparticles was investigated using bulk Al in a real industrial process using a commercial Al–Si hypoeutectic alloy. The modifier was prepared by hot extrusion approach. Its influence was investigated on Al and on commercial Al A356 alloy. The mechanical properties of the Al A356 alloy modified with WC nanoparticles was determined after T6 heat treatment and compared with an unmodified specimen of the same alloy. The results obtained in the modified Al A356 alloy reveal unusual behavior of the mechanical properties, where the elongation of the alloys improved by 32%–64%, while the tensile and the yield strengths remained unchanged. This behavior was attributed to a grain-size strengthening mechanism, where strengthening occurs due to the high concentration of grain boundaries, which prevent the dislocations motions in the metal lattice.

**INTRODUCTION**

The strengthening of metals is one of the main challenges in modern materials industry. Typically, such strengthening can be achieved by well-known alloying processes, in which some materials are added to affect the metallic product formed. The main disadvantage of the process is the relatively high fractions of expensive materials which are used to obtain the desirable technological characteristics.

Research on the formation of materials with improved mechanical performance is of great interest when they are considered in applications in the automotive or aerospace industry. In this work, the possibility of altering the mechanical properties and the microstructure of the Al–Si A356 casting alloy was investigated. Aluminum alloy A356 has excellent castability, mechanical characteristics, and physical properties. Its strength is usually improved by the alloying technique

[1–6] using master alloys, by heat treatment, or by applying ultrasound that affects the crystallization process of the solidified alloy [7–9]. Some studies have shown that the addition of $TiB_2$ particles, which serve as crystallization nuclei, causes refinement of the metal grains [10, 11]. Small additions of Sr and Na into Al–Si casting alloys enhance the formation of refined eutectic colonies [12, 13]. These grain-refinement processes cause improvement of the mechanical properties of metals. Another strengthening technology is Semi-Solid Metal processing (SSM), which was initially developed in the 70s of the last century [14]. Under the SSM approach, metals are treated by a mixture of fine solid nonmetal particles dispersed in the melt [15, 16].

In the last 20 years nanoscience approach has become a popular subject. It would be reasonable to expect that nanoparticles in the metal formed would cause such changes in the mechanical properties, but that their influence would be "milder" and more balanced. Unfortunately, the effect of additions of nanoparticles to metallic materials has not been thoroughly studied, but some researchers have reported that the mechanical properties of aluminum A356 could be improved by additions of $Al_2O_3$ or $TiB_2$ nanoparticles during the casting process [17, 18]. In the current work nanoparticles were used as nucleation sites; therefore, their melting point should be higher than that of aluminum to avoid their melting. Among their advantages, nanoparticles display high gas absorption on their surface because of the very large proportion of surface atoms out of the total number of atoms. Therefore, they need to be protected to prevent their floating on the surface of the molten metal.

WC used in this study as a modifier-based nanomaterial because of its advanced mechanical properties and relatively high melting point. Lekatou *et. al.* [19] showed that WC nanoparticles can be used in an aluminum reinforcement process in a relatively small amount of aluminum. Unfortunately, WC exhibits poor wettability by aluminum and its alloys. One solution was described by Chattopadhyay *et. al.* [20], who solved the problem by applying different treatment methods. In this work an alternative approach will be presented, where a modifier will be prepared by hot extrusion process.

It is relatively easy to perform the work and describe the experimental results obtained on relatively small amounts of the material, such as hundreds of grams. The aim of the present research is to investigate the influence of an addition of WC nanoparticles in a scaled-up process on the mechanical properties of the metal. The scaled-up industrial process is investigated on the commercial aluminum A356 casting alloy. An additional aim of the research is to describe the strengthening mechanism in the process.

**EXPERIMENTAL**

<u>Al–Si alloy sample preparation.</u>

WC nanoparticles (Inframat® Advanced Materials, 99.5%, crystallite size 40–70 nm) were mechanochemically activated with aluminum powder (Strem, 99%+, 20–40 mesh) in a Retsch PM 100 planetary ball mill. The milling speed was 400 rpm, and the activation time was 5 min. The powder mixture obtained was hot-extruded at 623K in a home-made extrusion container with an extrusion ratio equal to 17.4.

A 100 kg portion of ingots of Al–Si alloy A356 (Rheinfelden Alloys GmbH) was melted in an industrial electric resistance furnace, overheated to 1033 K, and then subjected to a standard industrial modification process using sodium and degassing by FDU. After degassing, a modifier containing 0.03 wt.% of WC nanoparticles (out of the total Al mass) was added to the molten metal, and the mixture was stirred for 10 min. The melt was poured into a special sand mold (Figure 1), whose central part solidifies much more slowly than the outer perimeter. The pouring temperature was controlled at about 1003 K. After pouring, the specimens produced were T6-heat-treated.

<u>Characterization techniques.</u>

Microstructure studies were carried out using a Zeiss Axiolab optical microscope. The specimens of the Al A356 casting alloy were analyzed using a Keller–Wilcox reagent (3ml HCl, 5ml $HNO_3$, 1ml HF, 190ml $H_2O$) to investigate their microstructures. Grain size measurements were made using the SIAMS 600 (System of Image Analysis and Modeling Structures) software version 2.0 and were based on the images taken from the optical microscope.

SEM images were obtained by a JEOL JSM-6510LV scanning electron microscope. The phases were identified by XRD analysis using a Panalytical X'Pert Pro X-ray powder diffractometer at 40 kV and 40 mA. The XRD patterns were recorded in the 2Θ range from 20º to 100º (the step size was 0.03º, and the time per step was 3 s). The chemical composition was detected by a Spectromax optical emission spectrometer.

The mechanical properties were measured by an Instron 3369 testing machine according to ASTM B 108-01 after T6 heat treatment. The modified alloys produced were compared to the unmodified alloys cast under the same conditions. Hardness tests were conducted on as-cast alloys by a Rockwell hardness tester (Wilson Hardness Ltd.) before and after modification.

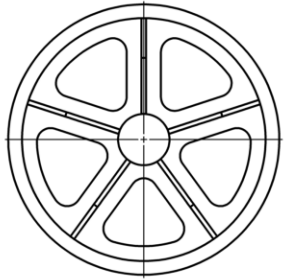

Figure 1: Special sand mold used in the research.

**RESULTS AND DISCUSSION**

Characterization of the modifier for Al-Si casting alloy.

The following experiments were conducted using the commercial A356 hypo-eutectic Al–Si casting alloy. Here the modifier was obtained by a hot extrusion process. During the process, the oxide layer on the nanoparticle surface disappeared because of the high compressive forces applied [21]. The modifier obtained is a cored wire filled with a mixture of a milled coarse Al powder with WC nanoparticles. An SEM image of the fiber structure obtained inside the modifier is presented in Figure 2.

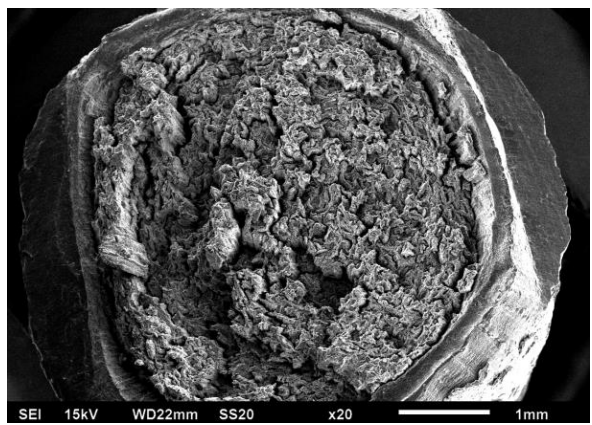

Figure 2. SEM image of a cored wire modifier.

Al-Si alloy casting process.

XRD patterns were used to reveal structural changes in the alloy during the modification process. XRD patterns of the modified and unmodified A356 casting alloys are shown in Figure 3.

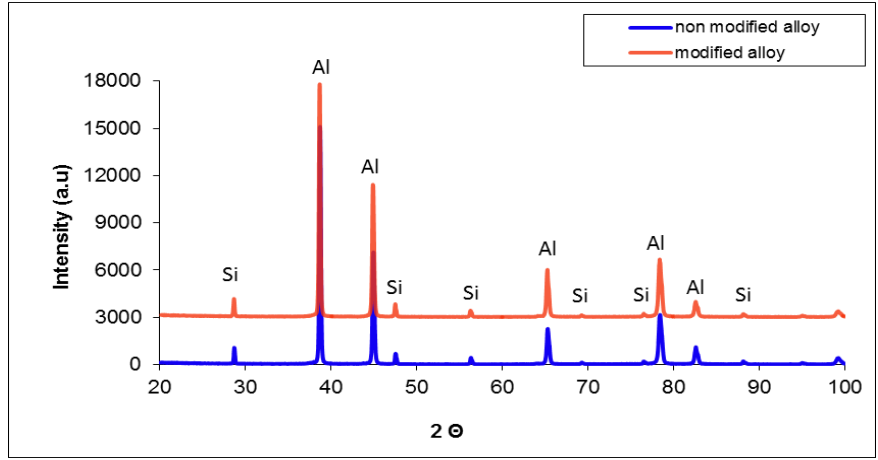

Figure 3. XRD patterns of aluminum alloy A356 before and after the modification process with the modifier based on WC nanoparticles.

The XRD results indicated the presence of only aluminum (JCPDS 01-071-4624) and silicon (JCPDS 01-070-5680) phases. No difference caused by the modification was observed; therefore, no new phase was formed in a detectable amount as a result of the modification process. These results were also confirmed by chemical analysis of the modified and unmodified alloys. The chemical compositions of both alloys determined by optical emission spectroscopy are presented in Table 1.

Table 1. Chemical composition of aluminum alloy A356 before and after the modification process with the modifier based on WC nanoparticles.

| State of the cast alloy | Si | Mg | Ti | Fe | Cu | Al |
|---|---|---|---|---|---|---|
| Before modification | 7.02 | 0.39 | 0.17 | 0.12 | 0.003 | Bal. |
| After modification | 7.14 | 0.37 | 0.17 | 0.12 | 0.003 | Bal. |

After T6 heat treatment, two specimens cut from the center and two from the perimeter of three different modified casting parts were subjected to tensile strength tests and compared with two specimens from the center and from the perimeter of one as-cast part. The crystallization rate on the perimeter of the cast part was much higher than the crystallization rate in the central area because of the different thicknesses of the sand mold.

The values of the tensile strength, yield strength, and elongation of the as-cast and modified alloys are summarized in Table 2.

Table 2. Mechanical properties of aluminum alloy A356 before and after the modification process with the modifier based on WC nanoparticles.

|  | State of the cast alloy | Tensile Strength [MPa] | Yield Strength [MPa] | Elongation [%] |
|---|---|---|---|---|
| Cast part center | Before modification | 277.5±1.1 | 226.7±1.8 | 2.8±0.1 |
|  | After modification | 284.8±0.9 | 226.7±1.6 | 4.6±0.3 |
| Cast part perimeter | Before modification | 303.5±13.1 | 231.8±3.6 | 6.5±1.8 |
|  | After modification | 307.3±7.6 | 230.3±4.4 | 8.6±0.8 |

It is seen from the table that the addition of the modifier based on WC nanoparticles improved the elongation of the metal by 32% and 64% in the perimeter and in the central part of the mold, respectively. The tensile strength and yield strength of the metal remained unchanged.

Generally, increasing the tensile strength reduces the ductility and *vice versa*. The results in Table 3 show unusual behavior, under which the elongation increases while the strength remains unchanged. The phenomenon of increases in the elongation as the grain size decreases is still not well understood. This phenomenon was also described by Li and Gui [22].

Hardness tests were conducted on the as-cast alloys before and after modification. The results are averages of three points in different regions of the cast ingots. The hardness of the as-cast alloy before modification was 14±3.5 HRB, and the hardness of the modified alloy was 23.3±2.1 HRB. The addition of WC nanoparticles causes the hardness to improve by 66%.

Hardness indicates the resistance of the material to localized plastic deformation and can indirectly indicate the tensile strength. In our case, it is valuable more as an indication of the resistance to penetration of the surface than as an indication of the tensile strength; therefore, its significant improvement was not expressed in the values of the ultimate tensile strength (UTS).

Specimens of the Al A356 casting alloy before and after modification were analyzed, and their microstructures are shown in Figure 4.

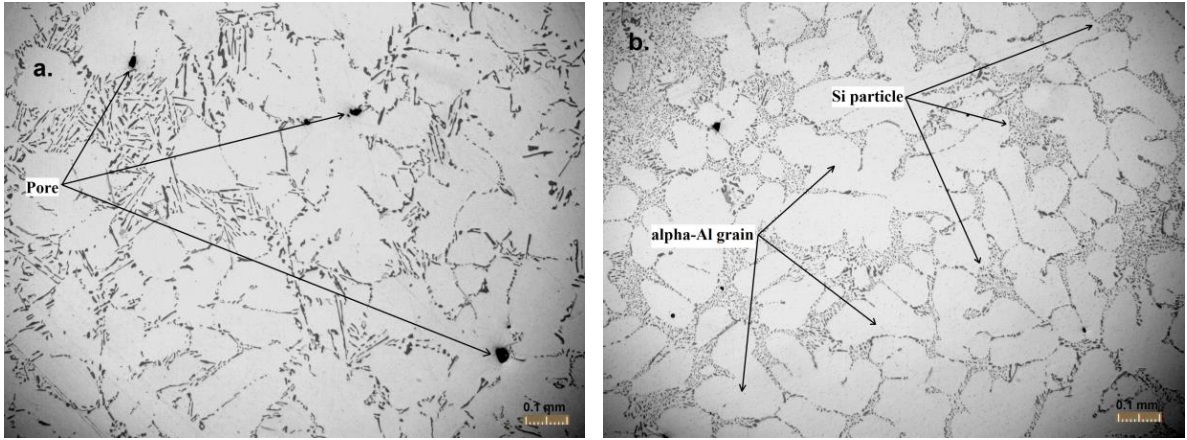

Figure 4. Optical micrographs of Al A356 alloy before (a) and after the modification process with the modifier based on WC nanoparticles (b) at 100× magnification.

It is seen from the micrographs that the microstructure of the modified alloy became finer and the number of pores (black dots on the microstructures) decreased. It is also seen that the modification process significantly refined the coarse large, elongated α-Al grains into fine equiaxed grains with homogeneously distributed eutectic Si particles.

The results of calculations of the average grain size using the SIAMS image analysis software are shown in Table 3.

Table 3. Calculated average grain size of A356 alloy before and after the modification process with the modifier based on WC nanoparticles.

| State of the alloy | Analyzed area (mm$^2$) | N$_o$ of grains on 1cm$^2$ | Average grain size (mm) |
|---|---|---|---|
| Before modification | 53.5 | 370 | 0.520 |
| After modification | 56.4 | 431 | 0.482 |

According to the calculated results, the average grain size of aluminum alloy A356 was reduced by 7.3% after modification with WC nanoparticles.

In summary, the modification of Al–Si casting alloy with WC nanoparticles is an important achievement for the technological casting process because of the improvement in the mechanical properties obtained without using common alloying compounds, which makes casting production more economically beneficial.

Strengthening mechanism.

The experimental results from the present research demonstrated a 66% improvement in the hardness of the Al–Si alloy. It is important to mention that hardness is a mechanical property which can indirectly indicate the tensile strength. In our case, it is more valuable for the resistance to penetration of the surface; therefore, its significant improvement was not expressed in terms of tensile strength values.

Usually the strengthening of aluminum can be increased by several mechanisms:

1) solid-solution strengthening when an alloying element forms a solid solution with the metal so that the atoms of the solute prevent dislocation motion in the lattice of the solvent;

2) grain-size strengthening expressed by the Hall–Petch relation [23], which calls for an increase in the yield stress of a polycrystalline material as its grain size decreases;

3) precipitation hardening, which is the process of high-temperature fixation of the structure of alloys at lower temperatures.

In our previous work we described the strengthening mechanism for an Al–Si alloy modified by ceramic nanoparticles [21]. Large concentrations of dislocations were observed near the grain boundaries, and nanoparticles were found only inside aluminum alloy grains. This is an indication that the nanoparticles act as nucleation accelerators in the crystallization process. As a result, the large numbers of grain boundaries formed prevent the slip motion of dislocations from one grain into another because of the lattice orientation difference.

According to the results obtained in the current research, together with our previous experimental results, it may be assumed that a grain boundary mechanism is responsible for the strengthening process.

## CONCLUSIONS

The effect of modification with WC nanoparticles on the microstructure and mechanical properties of the Al A356 casting alloy has been described in this work. Preparation of WC nanoparticles with the aluminum mixture followed by a hot extrusion process has been found to be a feasible method for preparing the Al modifier.

The commercial Al A356 hypoeutectic alloy was modified and investigated. Its elongation increased by 32% and 64% in different parts of the mold, and its hardness increased by 66%. Based on the previous experimental results together with the present research, it may be assumed that a grain-size strengthening mechanism operates in the modification of the Al–Si alloy with WC nanoparticles.

## ACKNOWLEDGMENTS

The authors would like to thank Dr. Alexey Kossenko, Mr. Alexander Krasnopolski and Ms. Natalia Litvak for their valuable help and support.

## REFERENCES

[1] Y. Birol: *J. Alloys Compd.*, 2012, 28(4), pp. 385-389.
[2] P.L. Schaffer and A.K. Dahle: *Mater. Sci. Eng. A*, 2005, 413-414, pp. 373-378.
[3] P.S. Mohanty and J.E. Gruzleski: *Acta Metall. Mater.*, 1995, 43 (5), pp. 2001-2012.
[4] C. Wang, M. Wang, B. Yu, D. Chen, P. Qin, M. Feng, and Q. Dai: *Mater. Sci. Eng. A*, 2007, 459, pp. 238-243.
[5] A. Daoud and M. Abo-Elkhar: *J. Mater. Proces. Tech.*, 2002, 120, pp. 296-302.
[6] Shang-Nan Chou, Jow-Lay Huang, Ding-Fwu Lii, and Horng-Hwa Lu: *J. Alloys Compd.*, 2006, 419, pp. 98-102.
[7] Y. Han, K. Le, J. Wang, D. Shu, and B. Sun: *Mater. Sci. Eng. A*, 2005, 405, pp. 306-312.
[8] A. Das and H.R. Kotadia: *Mater. Chem. Phys.*, 2011, 125, pp. 853-859.
[9] S. Zhang, Y. Zhao, X. Cheng, G. Chen, and Q. Dai: *J. Alloys Compd.*, 2009, 470, pp. 168-172.
[10] H.T. Lu, L.C. Wang, and S.K. Kung: *J. Chinese Foundrymen's Association*, 1981, 29, pp. 10-18.
[11] G.K. Sigworth and M.M Guzowski: *ASF Transactions*, 1985, 93, pp. 907-912.
[12] L. Clapham and R.W. Smith: *J. Crys. Growth*, 1986, 79 (1-3) part 2, pp. 866-73.
[13] S.A. Kori, B.S. Murty, and M. Chakraborty: *Mater. Sci. Eng. A*, 2000, A283, pp. 94-104.
[14] M.C. Flemings, R.G. Riek, and K.P. Young: *Mater. Sci. Eng.*, 1976, 25, pp. 103-117.

[15] P. Kapranos, P.J. Ward, H.V. Atkinson, and D.H. Kirkwood: *Mater. Design*, 2000, 21, pp. 387-394.
[16] Liao BoChao, Park Young Koo, and Ding HongSheng: *Mater. Sci. Eng. A*, 2011, 528 (3), pp. 986-995.
[17] S.A. Sajjadi, M. Torabi Parizi, H.R. Ezatpour, and A. Sedghic: *J. Alloys Compd.*, 2012, 511, pp. 226-231.
[18] M. Karbalaei Akbari, H.R. Baharvandi, K. Shirvanimoghaddam, Tensile and fracture behavior of nano/micro TiB2 particle reinforced casting A356 aluminum alloy composites, Materials and Design 66 (2015) 150–161.
[19] Lekatou , A.E. Karantzalis, A. Evangelou, V. Gousia, G. Kaptay, Z. Gácsi, P. Baumli, A. Simon: Aluminium reinforced by WC and TiC nanoparticles (ex-situ) and aluminide particles (in-situ): Microstructure, wear and corrosion behaviour, Materials and Design 65 (2015) 1121–1135.
[20] A.K. Chattopadhyay, P. Roy, S.K. Sarangi Study of wettability test of pure aluminum against uncoated and coated carbide inserts, Surface & Coatings Technology 204 (2009) 410–417.
[21] K. Borodianskiy, A. Kossenko, M. Zinigrad, *Improvement of the Mechanical Properties of Al-Si Alloys by TiC Nanoparticles*, Metall. and Mat. Trans. A vol. 44A (8) (2013) 4948-4953.
[22] S.X. Li and G.R. Gui: *J. Appl. Physics*, 2007, 101 (8), pp. 83525-83530.
[23] E.O. Hall: *Proc. Phys. Soc. B*, 1951, 64, pp. 747-753.